\documentclass[twocolumn]{article}

\usepackage[english]{babel}

\usepackage[letterpaper,top=2cm,bottom=2cm,left=3cm,right=3cm,marginparwidth=1.75cm]{geometry}
\usepackage[english]{babel}
\usepackage{amsmath}
\usepackage{graphicx}
\usepackage{authblk}
\usepackage[colorlinks=true, allcolors=blue]{hyperref}
\usepackage[utf8]{inputenc}
\usepackage[english]{babel}
\usepackage{csquotes}
\usepackage{natbib}
\usepackage{balance}
\usepackage{array}
\usepackage{booktabs}
\usepackage{tabularx} 
\usepackage{multirow}
\usepackage{adjustbox}

\title{What is Reproducibility in Artificial Intelligence and Machine Learning Research?}
\author[1]{Abhyuday Desai}
\author[1]{Mohamed Abdelhamid}
\author[2]{Nakul R. Padalkar}
\affil[1]{Ready Tensor, Inc.}
\affil[2]{Administrative Sciences, Metropolitan College, Boston University}

\usepackage{titlesec}
\titleformat{\section}
  {\normalfont\Large\bfseries\raggedright}{\thesection}{1em}{}

\titleformat{\subsection}
  {\normalfont\large\bfseries\raggedright} 
  {\thesubsection}  
  {1em}  
  {}  
\begin{document}

\twocolumn[
\date{}
  \maketitle
  \begin{abstract}
In the rapidly evolving fields of Artificial Intelligence (AI) and Machine Learning (ML), the reproducibility crisis underscores the urgent need for clear validation methodologies to maintain scientific integrity and encourage advancement. The crisis is compounded by the prevalent confusion over validation terminology. In response to this challenge, we introduce a framework that clarifies the roles and definitions of key validation efforts: repeatability, dependent and independent reproducibility, and direct and conceptual replicability. This structured framework aims to provide AI/ML researchers with the necessary clarity on these essential concepts, facilitating the appropriate design, conduct, and interpretation of validation studies. By articulating the nuances and specific roles of each type of validation study, we aim to enhance the reliability and trustworthiness of research findings and support the community's efforts to address reproducibility challenges effectively.
\end{abstract}
  \vspace{1cm}
]

\begin{figure*}[t]
  \centering
\includegraphics[width=1.0\linewidth]{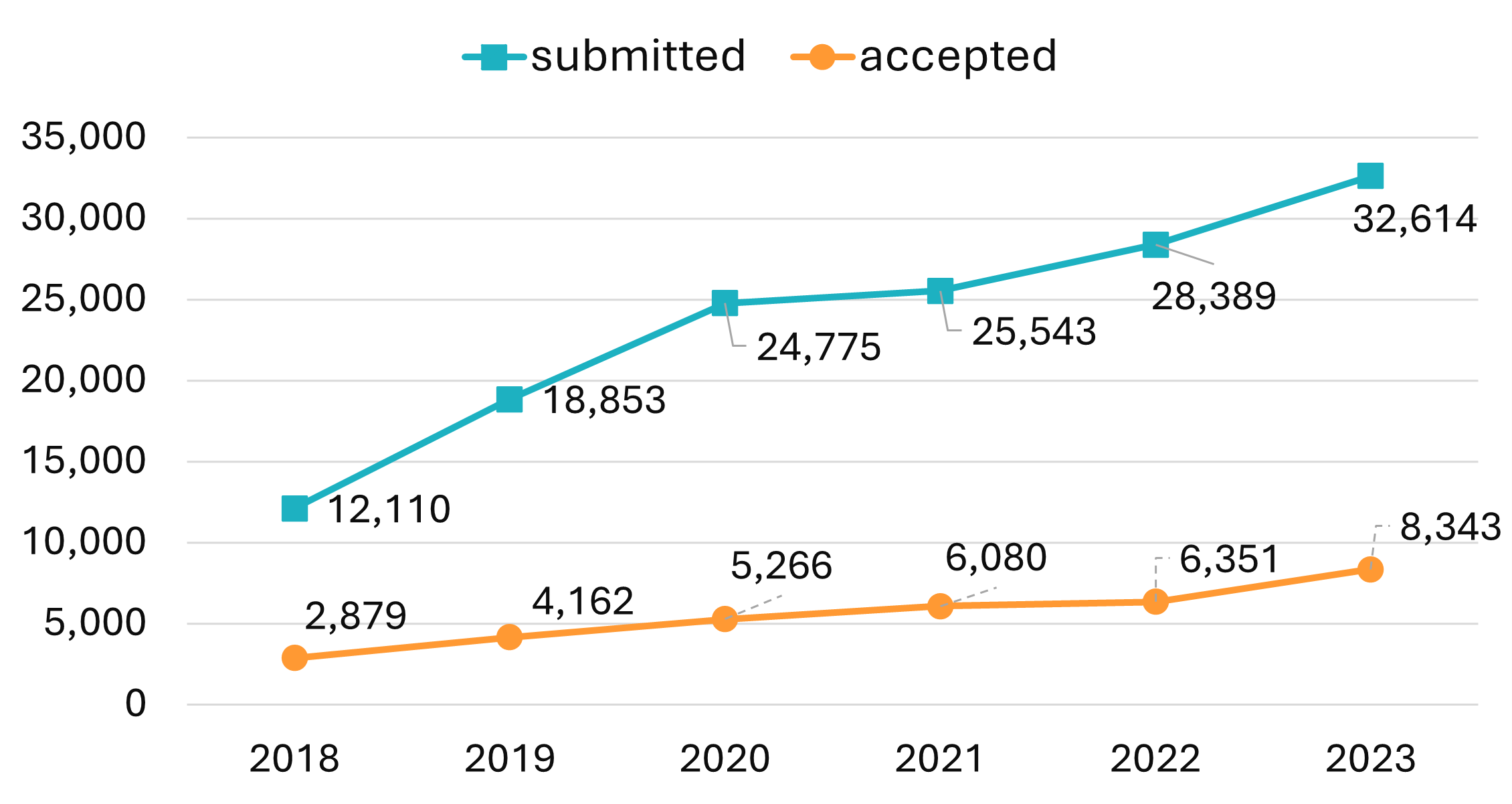}
\caption{\label{fig:publications_chart_}Total number of papers submitted and accepted at NeurIPS, ICML, ICLR and AAAI from 2018 to 2023}
\end{figure*}

\section{Introduction}

The AI/ML domain has witnessed explosive growth in research publications over the past few years. With major conferences receiving thousands of submissions annually, the sheer volume of research output has made it challenging to ensure consistent reproducibility. Conferences like NeurIPS, ICML, ICLR, and AAAI have seen a significant growth in paper submissions. The number of papers submitted to these four conferences has increased by 169\% between 2018 and 2023. Figure ~\ref{fig:publications_chart_} shows the number of articles submitted and accepted at the top conferences.

Although the number of publications has increased, attempts to reproduce the findings of these papers have often been met with challenges. Numerous studies have highlighted this critical issue in recent years. For example, a survey of 1,576 researchers in the Nature journal indicated that more than 70\% of researchers have tried and failed in their attempt to reproduce another researcher’s experiments, and more than half have failed to reproduce their own experiments \citep{Baker2016-BAKSL-2}.

\cite{collberg2016repeatability} attempted to execute the code from 601 papers from computer systems research. This study only involved attempts to execute the code, not verify the correctness of the published results. These attempts to re-execute the code were divided into three categories: category 1: time to examine a research artifact is limited and communicating with the author is not an option; category 2 involves the scenario where ample time is available, but the lead author is not available for consultation; category 3 represents the case where ample time is available, and the author is available to correspond.  They achieved success in 32.3\% in category 1 attempts, 48.3\% in category 2, and 54.0\% in category 3.

\cite{raff2019step} attempted to reproduce results of 255 papers between 1984 and 2017 with a success rate of 63.5\%. This study involved independent reproduction attempts, where the original authors’ code, even if available, was not used.

\cite{islam2017reproducibility} investigated the reproducibility of benchmarked deep reinforcement learning tasks and found a wide range of results reported in the literature for the same baseline algorithms, highlighting the difficulty of reproduction attempts.  These variations were attributed to factors such as external randomness, under-reporting of hyperparameters, and a narrow range of tasks under the benchmark.

\cite{pham2020problems} revealed a striking statistic that highlights reproducibility challenges in deep learning research. They reported that for 16 identical training runs for a popular deep learning network architecture called LeNet5, the accuracy of the resulting 16 models ranged from 8.6\% to 99.0\% - a difference of 90.4\% across the runs.

The reproducibility challenges in AI/ML research highlighted above, spanning from code execution issues to the variability in experimental outcomes, emphasize the critical need for a systematic approach to ensure effective and reliable validation efforts. 

\section{Terminology Confusion}
Addressing the reproducibility crisis in AI and ML research requires not only clear validation methods but also a precise understanding of fundamental terms, like 'repeatability', 'reproducibility', and 'replicability'. Despite their significance in establishing the trustworthiness of research, there exists a notable confusion over these terms' meanings. For instance, \cite{hunold2015survey} highlighted findings from a survey conducted at the Euro-Par conference on reproducibility in parallel computing, where only 32\% of respondents indicated they could accurately differentiate between replicability, repeatability, and reproducibility, pointing to a significant gap in understanding that complicates the validation process.

\cite{gundersen2021fundamental} highlights this issue further by reviewing the 34 diverse definitions and interpretations of 'reproducibility', ‘replication’, and related terms across numerous research papers, concluding that a single, agreed-upon definition does not exist. \cite{barba2018terminologies} work on the disparities in terminologies for reproducible research reveals how the application and understanding of these terms can vary significantly across researchers in different scientific fields, contributing to ongoing debates and misunderstandings. Barba’s study identifies three distinct approaches for using these terms: 

\begin{itemize}
  \item Approach A: No distinction is made between 'reproducibility,' 'replicability,' and 'repeatability.' 
  \item Approach B1: 'Reproducibility' is defined as using the original data and code to regenerate the results, whereas 'replicability' refers to generating similar scientific findings with new data.
  \item Approach B2: Conversely, 'reproducibility' implies that independent researchers achieve the same results using their own methods and data, while 'replicability' involves using the original study's artifacts.
\end{itemize}

The most prominent example of category A is the Open Science \cite{open2012open}, an initiative focusing on promoting open science practices across disciplines. Here is an excerpt from chapter 11: “narrowly, reproducibility is the repetition of a simulation or data analysis of existing data by re-executing a program. More broadly, reproducibility refers to direct replication, an attempt to replicate the original observation using the same methods of a previous investigation but collecting new [data].”

Under category B1, there are often-cited works by \cite{claerbout1992electronic}, \cite{buckheit1995wavelab}; are credited for being pioneering works in defining reproducibility. While \cite{peng2006reproducible} is cited for distinguishing the term replication. The works under category B2 include \cite{cartwright1991replicability}, \cite{drummond2009replicability}, and \cite{plesser2018reproducibility}. These researchers directly swapped the meaning of the terms reproducibility and replicability compared to Category B1. 

There have been attempts to resolve the debate on terminology. \cite{patil2016} provide their definition of reproducibility and different forms of replicability. However, they encountered conflicts with their definition of reproducibility involving the case where the data and code from the original study were incomplete and/or incorrect. The reproducibility study team implemented new code, which indeed reproduced the original results. The authors noted that this didn’t meet their technical definition of reproducibility. 

\cite{goodman2016does} presents a new lexicon for research reproducibility, introducing terms like methods reproducibility, results reproducibility, and inferential reproducibility. These terms avoid the ambiguity resulting from the different interpretations of the words “reproducible”, “replicable,” and “repeatable” in common usage. \cite{gundersen2021fundamental} also proposes new terms representing three degrees of reproducibility: outcome, analysis, and interpretation reproducibility. The authors also define four reproducibility types driven by documentation that is used to conduct the reproducibility study: R1 (only text is used as reference), R2 (text and code are used), R3 (text and data are used), and R4 (text, data, and code are used). 

\section{Motivation for Our Framework}

The terminological confusion in reproducibility research hampers scientific progress. Rather than introducing new terms, our framework refines existing ones to establish a common language that researchers already intuitively understand. 

Current approaches to reproducibility often emphasize convenience by sharing code and data to facilitate regeneration. While transparency through artifact sharing is valuable, it does not inherently enhance scientific trust. True validation occurs when findings persist despite deliberate variations in experimental elements.

We propose a hierarchical validation framework that places different types of validation efforts on a spectrum of increasing scientific rigor. While repeatability (obtaining consistent results under identical conditions) forms the baseline, it represents only a minimal validation standard. Reproducibility, which involves external researchers validating the original experiment's correctness, elevates this standard further.

The highest levels of validation come through replication efforts: direct replicability tests whether findings hold when implementation details change, while conceptual replicability examines whether conclusions remain valid across entirely different experimental approaches.

This hierarchy challenges the research community to move beyond making repetition more convenient toward facilitating higher forms of validation. It also provides clear guidance when validation studies yield conflicting results: findings from higher-level validation efforts should generally take precedence over those from lower levels. This precedence relationship, often missing from existing frameworks, offers researchers a clearer path through the complex landscape of validation efforts.

\section{AI/ML Research Context and Reproducibility Landscape}

\subsection{Scope and Definitions: AI and ML in Context}

To establish a common understanding and provide context for the discussion on reproducibility, we first define the key terms relevant to this paper: Artificial Intelligence (AI), Machine Learning (ML), Supervised Learning, Unsupervised Learning, Reinforcement Learning, and Generative AI. These definitions help clarify the scope of our framework, especially as each paradigm presents unique characteristics and requirements.

\textbf{Artificial Intelligence (AI)}: refers to computer systems capable of performing tasks that typically require human intelligence. 

\textbf{Machine Learning (ML)}: is a subset of AI that focuses on systems that learn from data without explicit programming, improving performance through experience.

Our paper focuses on empirical, data-centric AI/ML research, where the central goal is to investigate how models learn from data to perform tasks. This encompasses most contemporary AI/ML research published in venues such as the Conference on Neural Information Processing Systems (NeurIPS), the International Conference on Machine Learning (ICML), the International Conference on Learning Representations (ICLR), and the AAAI Conference on Artificial Intelligence (AAAI). 

Within the empirical domain, several paradigms have emerged, each with distinct validation challenges. These paradigms represent different approaches to leveraging data for building intelligent systems, and our framework aims to provide validation guidance across this spectrum.

\subsubsection{Machine Learning}

\textbf{Supervised Learning} involves training on labeled data pairs. Common reproducibility issues include data leakage (i.e. unintentional inclusion of test data during training), inconsistent data-splitting protocols, unreported preprocessing steps, and variations in cross-validation implementations—all of which can significantly impact model performance.

\textbf{Unsupervised Learning} discovers patterns in unlabeled data. Reproducibility challenges stem from sensitivity to initialization methods, ambiguous convergence criteria, inconsistently applied evaluation metrics, and unreported parameter choices that can lead to substantially different discovered patterns.

\textbf{Reinforcement Learning (RL)} trains agents through interaction and feedback from the environment. As noted in \cite{islam2017reproducibility}, the reproducibility of RL is complicated by environmental stochasticity, variations in the implementation of reward functions, inconsistent episode termination conditions, and differences in exploration strategies - factors often not adequately documented in research articles.

\subsubsection{Deep Learning}
Deep learning, a subset of machine learning, has powered breakthroughs in computer vision, natural language processing, by leveraging multi-layer neural networks. Reproducibility challenges include sensitivity to weight initialization schemes, dependencies on specific hardware/software configurations, undocumented hyper-parameter tuning protocols, and random seed dependencies. For example, \cite{pham2020problems} reported drastically different performance outcomes (8.6\% to 99.0\%) when training LeNet5 under identical hyperparameters but with different random seeds.

\subsubsection{Foundation Models and Generative AI}
Foundation Models and Generative AI are further subsets within deep learning. Foundation models (e.g., large language models) are trained on vast datasets using substantial computational resources. Reproducing or fully retraining such models is often infeasible, leading researchers to rely on downstream evaluations or partial re-implementations for validation. Reproducibility challenges include incomplete documentation of training data, limited access to computational infrastructure, and difficulties in accurately comparing different implementations of the same model architecture. As noted by \cite{miikkulainen2024generative}, these models represent a significant shift in AI development that challenges traditional evaluation practices, requiring new approaches to validation. 

\subsection{Reproducibility Landscape: From Concepts to Implementation}
The reproducibility landscape in AI/ML research can be categorized into conceptual frameworks, standards/certification systems, and practical tools/platforms. Understanding these categories helps position our contribution within existing efforts.

\subsubsection{Conceptual Frameworks and Definitions}
At the foundational level, researchers have proposed various conceptual frameworks defining what reproducibility means. For example, \cite{goodman2016does} proposed "methods reproducibility," "results reproducibility," and "inferential reproducibility" to distinguish different aspects of validation. Similarly, \cite{gundersen2021fundamental} offered a taxonomy based on documentation types (R1-R4) and degrees of reproducibility (outcome, analysis, interpretation). These conceptual frameworks provide the theoretical foundation for reproducibility efforts. The proposed reproducibility framework falls under this category. 

\subsubsection{Standards and Certification Systems}

Building on conceptual foundations, standards and certification systems provide operational criteria for evaluating reproducibility. \cite{acm_artifact_badging}
introduced a badging system to certify research artifacts at different levels, from "Available" to "Evaluated" to "Reproduced." with the goal to promote research validation and integrity. \cite{heil2021reproducibility} established bronze, silver, and gold standards focusing on increasing levels of automation in reproduction. 

Major conferences including \cite{neurips2021checklist} and \cite{icml2024guidelines} have implemented paper checklists requiring documentation of key experimental details to address issues of reproducibility, transparency, ethics, and societal impact. \cite{iclr2022authorguide} strongly encourages authors to include a Reproducibility Statement within their paper to explain efforts made to ensure reproducibility. 

\cite{kapoor2023leakage} developed model information sheets standardizing documentation for ML models to identify and prevent data leakage issues, while \cite{fursin2021collective} created the Collective Knowledge framework establishing standards for modular component-based reproducibility. 

\subsubsection{Implementation Tools and Platforms}
The third category comprises practical tools and platforms supporting reproducible practices. For example, \cite{paperswithcode} links research papers to implementation code, creating verifiable connections between claims and implementations. Other platforms like \cite{wandb} and \cite{neptune} provide experiment tracking infrastructure that automatically documents research artifacts.

\section{The Reproducibility Framework}

\begin{figure*}[!t]
  \centering
\includegraphics[width=1.0\linewidth]{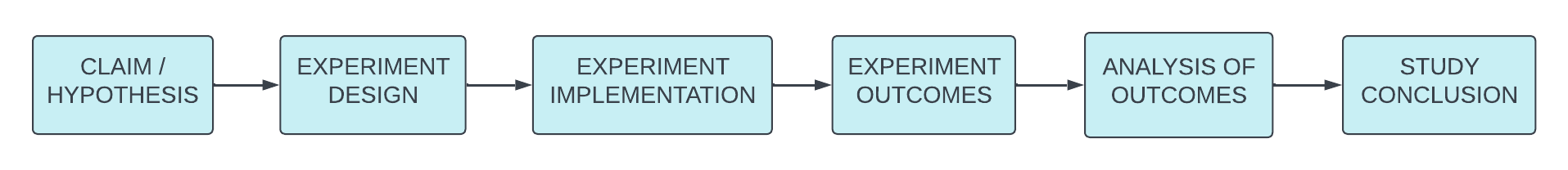}
\caption{\label{fig:workflow}The research publication workflow}
\end{figure*}

\subsection{Components of Research Studies}
\cite{gundersen2021fundamental} proposes his definitions for reproducibility terms by referring to the general scientific method in research studies. We use Gundersen’s approach while narrowing the focus on the key components incorporated in a research publication. 

A research publication consists of the components, as visualized in Figure \ref{fig:workflow}:

\begin{enumerate}
  \item \textbf{Claim/hypothesis:} Represents the central assertion or prediction that the study aims to investigate. It is the result that the authors are trying to prove. 
  
  \item \textbf{Experiment design:} This is the structured plan devised to test the claim or hypothesis. It encompasses the selection of methodologies for data collection, data pre-processing, model building, and validation. This stage establishes the research protocol, detailing how the experiment will be conducted to ensure valid and reliable results.
  
  \item \textbf{Experiment implementation:} Refers to the practical execution of the experiment according to the predefined design. This phase involves the operational aspects of the study, such as collecting data, coding and scripting for analysis or model building, and utilizing necessary hardware or software resources.

  \item \textbf{Experiment outcomes:} These are the raw results obtained from the experiment. These outcomes are the direct observations or data points collected during the implementation phase, before any analysis or interpretation.

  \item \textbf{Analysis of outcomes:} This is the step of examining and interpreting the experiment outcomes through various analytical techniques. This may include statistical testing, the creation of summary tables and charts, and other methods to extract insights and understand the data's implications regarding the hypothesis.

  \item \textbf{Study conclusion:} Refers to the final assessment of the hypothesis based on the analyzed outcomes. This stage involves synthesizing the findings to determine whether the initial claim is supported or refuted by the experimental evidence, culminating in a clear statement about the validity of the claim.
\end{enumerate}

\subsection{The Validation Spectrum}
We now introduce our reproducibility framework by defining key types of validation studies summarized in Table \ref{tab:table}. 

\renewcommand{\arraystretch}{1.3}

\begin{table*}[ht!]
\centering
\caption{Defining types of validation studies}
\label{tab:table}
\centering
\newcolumntype{A}{>{\hsize=0.60\hsize\centering\arraybackslash}X}
\newcolumntype{B}{>{\hsize=0.55\hsize\centering\arraybackslash}X}
\newcolumntype{C}{>{\hsize=0.52\hsize\centering\arraybackslash}X}
\makebox[\linewidth][c]{
  \small
  \begin{tabularx}{1.19\linewidth}{|A|B|B|B|B|B|C|C|}
  \hline
   & \textbf{Researchers} & \textbf{Claim / Hypothesis} & \textbf{Experiment Design} & \textbf{Experiment Implementation} & \textbf{Experiment Outcomes} & \textbf{Analysis of Outcomes}  & \textbf{Conclusions} \\
  \hline
  Repeatability & same & same & same & same & same & same &same \\
  \hline
  Reproducibility & different & same & same & same & same & same & same \\
  \hline
  Direct Replicability & different & same & same & different & different & same / different & same \\
  \hline
  Conceptual Replicability & different & same & different & different & different & same / different & same\\
  \hline
  \end{tabularx}
}
\end{table*}

\textbf{Repeatability}
Repeatability refers to the ability to obtain consistent results by the same team using the same experimental setup, including the claim or hypothesis, experiment design, implementation, outcomes, analysis of outcomes, and conclusions. This means that when the original researchers re-execute their experiment under the same conditions, they achieve the same findings.

\textbf{Reproducibility:}
Reproducibility involves external researchers validating the correctness of an original experiment’s findings by following the documented experimental setup. This can be achieved through direct use of the original data and code, called dependent reproducibility, or by reimplementing the experiment, titled independent reproducibility, ensuring the results are reliable and applicable across different teams. Importantly, validating the correctness of the implementation in both dependent and independent reproducibility distinguishes these efforts from mere repeatability, enhancing the study's scientific rigor.

\textbf{Direct Replicability:}
Direct Replicability is when an independent team intentionally varies the implementation of an experiment—while keeping the hypothesis and experimental design consistent with the original study—to verify the results. This deliberate alteration may entail using different datasets, methodologies, or analytical approaches. The objective is to affirm the original findings under slightly altered experimental conditions, but not entirely new ones, ensuring that the study's conclusions are not solely contingent on the original set of experimental parameters.

\textbf{Conceptual Replicability:}
Conceptual Replicability refers to the process where an independent team tests the same hypothesis through a fundamentally new experimental approach. Unlike direct replicability, which alters only the implementation, conceptual replicability involves redesigning the experimental setup itself. This approach aims to validate the hypothesis in broader or different contexts by significantly deviating from the original study's design. 
Let us now explore these validation types in more detail.

\subsection{The Role of Repeatability}
Repeatability is the exercise of ensuring that results are consistently achievable under the same experimental conditions by the original researchers. When different researchers repeat the experiment without verifying its correctness, we term this \textbf{external repeatability}.

External repeatability, while extending the concept to include independent researchers, does not materially enhance the trustworthiness or reliability of the findings beyond what is achieved by the original team's repetition. The absence of a thorough examination of the experiment’s design or its outcomes means that this effort remains a basic repetition rather than a deepened form of validation.

\subsection{Reproducibility for Validating Experiment Correctness}
Reproducibility transcends the mere blind regeneration of results; it fundamentally involves a critical validation of the original study's implementation correctness. This process ensures that, when an experiment implementation is as described and without material flaws, it consistently leads to the same conclusions, affirming the reliability of the original findings.\newline

\textbf{Clarifying the Scope of Reproducibility}\par
Reproducibility focuses on validating the correct implementation of the experiment design and its ability to consistently lead to the original study's conclusions. It does not assess whether the experimental design is the most appropriate or the only option for testing the hypothesis; this validation is categorized under replicability in our framework. \\ \\ 
\textbf{Pathways to Achieving Reproducibility}\par

There are two main pathways to pursue a reproducibility study:
\begin{enumerate}
    \item \textbf{Dependent Reproducibility:} This approach uses the original study's code and data to recreate the results and relies on the availability of these original artifacts.

    \item \textbf{Independent Reproducibility:} Entails a researcher independently reimplementing the experiment based on the published descriptions, without reliance on the original code and data.
\end{enumerate}

Independent reproduction stands as a more rigorous validation method compared to dependent reproduction due to its requirement for the reproducer to independently parse and execute the original experiment's design. This method deepens the validation process, making it more probable to identify flaws or limitations in the original findings than when relying on dependent reproduction. \\ \\
\textbf{Navigating Common Scenarios}\\
For further clarity around our definition of reproducibility, consider the following scenarios: \\ \newline
\textbf{Discovering Flaws in the Original Study:} \newline
Reproduction attempts of AI/ML research may reveal flaws in the original study, such as data leakage or insufficient control of experimental variables, affecting outcomes significantly. If corrections to these flaws change the study's conclusions, it indicates a failure in reproducibility, pointing to problems with the original implementation’s accuracy. This underscores that reproducibility goes beyond mere re-execution of an experiment by different individuals. It requires an independent verification of the experiment’s implementation for correctness.\newline

\textbf{Adherence to Experiment Methodology:} \newline
Reproducibility is achieved by closely following the original experiment's methodology, where minor modifications are permitted but should not be confused with direct replication. For instance, translating code from TensorFlow to PyTorch, while a technical modification, falls under the reproduction category if it follows the original experiment design and hypothesis testing. These minor changes do not constitute direct replication, which involves a deliberate alteration in the experiment’s implementation to test the robustness of findings under varied conditions.

\subsection{Replicability for Robustness and Generalization}

\begin{figure*}[t]
  \centering
\includegraphics[width=0.95\linewidth]{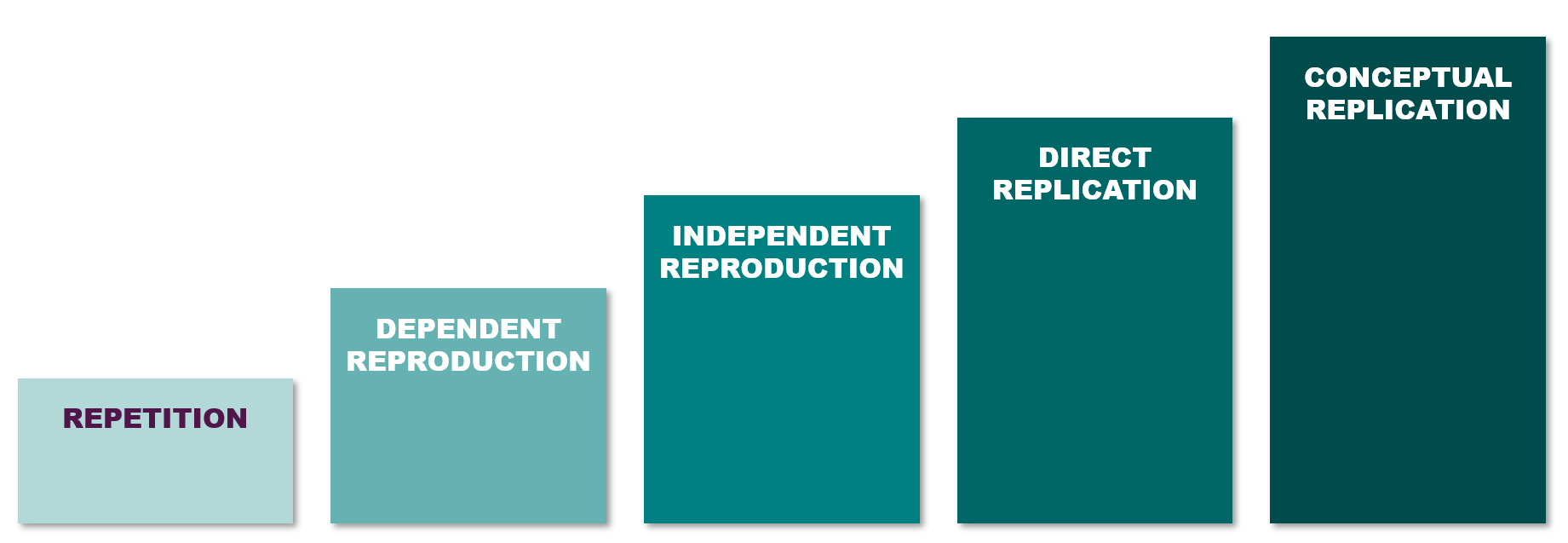}
\caption{\label{fig:validation_spectrum}Hierarchy of validation studies in research}
\end{figure*}

The overarching goal of replicability, encompassing both direct and conceptual approaches, is to validate the robustness and generalization of a study’s findings. By testing the original results through intentional alterations in experiment implementation, called direct replicability, or entirely new experimental designs, titled conceptual replicability, researchers aim to confirm that the conclusions drawn are not merely artifacts of specific experimental setups but hold across different contexts and approaches. Replicability can help understand the boundaries within which the original findings hold.

Direct replicability focuses on intentional modifications to an experiment’s implementation while adhering to its original design and hypothesis. These modifications may include using alternative datasets, analytical methods, computational tools, or statistical techniques, aiming to test the stability of the findings under slight but deliberate variations.

Conversely, conceptual replicability represents a more extensive departure from the original experiment by adopting entirely new designs to explore the same hypothesis. This approach allows for testing the hypothesis under significantly different conditions or assumptions, potentially uncovering new insights, or challenging the original study’s conclusions.

It is important to distinguish conceptual replicability from corroboration. While corroboration involves gathering supportive evidence from multiple studies to reinforce a theory, conceptual replicability specifically examines the hypothesis's validity across diverse experimental designs, without necessarily aiming to support the overarching theory directly. 

\subsection{Validation Hierarchy and Scientific Rigor}
Let us now explore the implications of the different types of validation studies on scientific rigor and reliability. As illustrated in Figure ~\ref{fig:validation_spectrum}, the validation hierarchy from repeatability to conceptual replicability represents a continuum of increasing scientific rigor and reliability. 

Repeatability establishes the baseline integrity of research findings, ensuring that results are not artifacts of chance. It is the essential first check for any study, affirming internal consistency.

Dependent Reproducibility and Independent Reproducibility elevate scrutiny by testing if the findings hold when the experiment is recreated. Dependent reproducibility involves using the original materials and validating the correctness of the implementation as described in the study. Independent reproducibility is achieved by reconstructing the experiment based on the original study’s methodology and similarly validating the implementation's correctness. Both approaches ensure that the experimental results are not only repeatable but also methodologically sound, reinforcing the reliability of the findings.

Direct and Conceptual Replicability represent further steps in validating research findings. Direct replicability tests the original conclusions against variations in methodology, probing the findings' stability across different implementations. Conceptual replicability, the pinnacle of validation efforts, explores the hypothesis in entirely new contexts, assessing the generalizability and versatility of the conclusions. 

This validation hierarchy illustrates that while repeatability and reproducibility are fundamental for trust in the original experiment’s findings, direct and conceptual replicability are crucial for demonstrating the findings' robustness and generalization to methodological changes. Importantly, each higher level of validation can supersede the previous level. For instance, if a study’s results are independently reproducible—even if not dependently reproducible—or directly replicable in the absence of the ability to reproduce using the original dataset, it underscores the reliability of the findings at a higher degree of scientific scrutiny. Each step increases the rigor and provides a deeper understanding of what the research achieves and its limitations in the face of expanding scientific scrutiny.

\section{Framework Analysis and Validation Through Case Studies}
Our framework both builds upon existing reproducibility approaches and provides novel contributions to address the terminological confusion in the field. This section analyzes how our framework relates to prominent existing approaches and presents case studies demonstrating its practical application.
\subsection{Comparative Analysis with Existing Frameworks}

\textbf{Gundersen's Outcome, Analysis, and Interpretation Reproducibility}: Gundersen's categorization focuses on \textit{which aspects} of a result are validated: outcomes (same results), analysis (same analysis despite different outcomes), or interpretation (same conclusions despite different outcomes and analyses). In contrast, our framework emphasizes \textit{how} validation is conducted—through repeatability, reproducibility, or replicability—regardless of which aspects are confirmed.

Integrating these approaches yields a more comprehensive evaluation system. For example, an independent reproduction study could achieve outcome reproducibility (identical results), analysis reproducibility (different outcomes, but same analysis), or inferential reproducibility (same conclusions through different outcomes and analysis). This integration clarifies both the validation method employed and the specific aspects confirmed, providing a richer description of scientific reliability.

\textbf{Gundersen's R1-R4 Documentation Levels}: Gundersen's R1-R4 levels categorize studies based on documentation completeness: R1 (description only), R2 (description and code), R3 (description and data), and R4 (description, data, and code). These levels determine what validation types are practically feasible. Gundersen's framework complements ours by highlighting the prerequisites for different validation types.

\textbf{Goodman's Reproducibility Lexicon}: Goodman's framework distinguishes methods reproducibility (same procedures), results reproducibility (same outcomes with new data), and inferential reproducibility (same conclusions). This parallels Gundersen's outcome-analysis-interpretation categorization but with different terminology. The key distinction is our emphasis on the hierarchy of validation rigor, establishing that conceptual replicability provides stronger evidence than direct replicability and dependent reproducibility surpasses repetition.

\textbf{Heil's Bronze, Silver, Gold Standards}: Heil's standards focus on automating reproduction: bronze (manual effort required), silver (semi-automated), and gold (fully automated one-click regeneration of results). A critical distinction in our framework is that repeatability, even when fully automated per Heil's gold standard, offers limited scientific assurance without thorough investigation of implementation correctness. The ease of regenerating results does not substitute for critical evaluation of methodological soundness.

\subsection{Reproducibility Framework in Practice}

This section presents case studies demonstrating validation attempts across our framework's hierarchy. These examples illustrate how different validation levels from repeatability to conceptual replicability provide increasing scientific evidence and reveal distinct insights, while adapting to diverse research contexts from traditional ML studies to foundation models where conceptual replicability may be the only viable approach. \\

\textbf{Example 1: From Repeatability to Reproducibility}

The critical distinction between repeatability and reproducibility is starkly illustrated by a high-profile study by \cite{just2017machine} that claimed to identify suicide risk from brain scans using machine learning. The paper reported achieving 91\% accuracy in identifying suicidal ideation from fMRI data, garnering significant media attention and clinical interest. The authors provided code and a portion of their study data for public access.

When \cite{verstynen2023overfitting} attempted to reproduce study findings, they discovered fundamental flaws in the methodology. Their examination revealed information leakage in the machine learning pipeline; the features of the classifier were effectively tuned to the particular dataset, with information from the holdout set inappropriately influencing the cross-validation process. This methodological flaw meant the reported 91\% accuracy was a substantial overestimation, eventually leading to the paper's retraction.

This type of data leakage is not an isolated problem. A comprehensive survey by \cite{kapoor2023leakage} identified 294 studies across 17 scientific fields affected by various forms of data leakage, often resulting in significantly overoptimistic conclusions. In a case study of civil war prediction, they found that when leakage errors were corrected, complex ML models showed no substantive performance advantage over decades-old logistic regression approaches—contradicting previously published findings.

This case exemplifies why reproducibility involves more than merely re-executing experiments. By critically examining the implementation's correctness, reproduction efforts revealed fatal methodological flaws that invalidated the study's conclusions. This underscores why our framework distinguishes between repeatability and reproducibility, with the latter providing substantially stronger scientific evidence through methodological validation.  \\

\textbf{Example 2: From Reproducibility to Replicability}

The distinction between reproducibility and direct replicability is clearly demonstrated in a study by \cite{zech2018variable} examining deep learning models for pneumonia detection. The researchers developed convolutional neural networks (CNNs) that achieved excellent performance (AUC up to 0.931) when validated on test sets from the same hospital systems used for training. However, when attempting direct replication by applying these models to other hospital systems with different demographic distributions, performance degraded significantly. 

Further investigation revealed the models were primarily detecting hospital-specific features rather than generalizable medical findings, with CNNs identifying the source hospital with 99.95\%+ accuracy. Differences in disease prevalence between institutions created confounding variables that the models exploited but failed to generalize across systems.

This case illustrates why validation must progress beyond reproducibility to replicability. While the pneumonia detection models satisfied reproducibility criteria by performing consistently within their original context, they failed at direct replicability when the implementation context changed. This limitation became apparent only through direct replication efforts, highlighting how our validation hierarchy provides increasingly valuable insights about the robustness and generalizability of AI/ML research by placing higher importance on direct and conceptual replicability.  \\

\textbf{Example 3: Conceptual Replicability Revealing Limitations}

The Synthetic Minority Over-sampling Technique (SMOTE) by \cite{chawla2002smote} provides an instructive example where successful replication studies refined understanding of a widely-adopted technique's limitations. SMOTE, which generates synthetic examples for minority classes to address class imbalance, gained widespread adoption after its introduction, with numerous studies reproducing its reported benefits across various classification tasks.

Conceptual replication studies, which tested the same hypothesis using fundamentally different experimental designs, revealed important nuances in SMOTE's effectiveness. For instance, \cite{van2007experimental} showed that simpler alternatives like random undersampling sometimes outperformed SMOTE, particularly for certain classifier types and evaluation metrics. A comprehensive study by \cite{abdelhamid2024balancing} discovered substantial variability in performance of class imbalance handling techniques across diverse datasets, emphasizing the need for dataset-specific analysis in choosing the best class imbalance handling technique. 

This case demonstrates how conceptual replication — the highest level in our validation hierarchy — can reveal nuanced understanding beyond what reproduction or direct replication might uncover. By testing the central hypothesis through diverse experimental approaches, these studies identified boundary conditions and comparative limitations that refined the research community's understanding of when and how to apply SMOTE effectively.  \\

\textbf{Example 4: Conceptual Replicability for Foundation Models}

Large Language Models (LLMs) and other generative AI systems present unique validation challenges that necessitate conceptual replicability as the primary validation approach. As \cite{miikkulainen2024generative} notes, these models represent a paradigm shift in AI development where "each such model is a one-off — not unlike a human that is shaped by their experience... We have to learn to do science with a sample of one."

The spectrum of model accessibility, from proprietary such as ChatGPT-4 from \cite{openaiChatGPT2023}, to open-weights such as Llama 3 by \cite{meta2024llama3}, to fully open-source, makes traditional reproducibility approaches often impractical or impossible. The extraordinary computational requirements, proprietary training data, and sheer complexity of these systems mean that even with complete documentation, dependent reproduction would remain infeasible for most research teams. For such cases, validation adapts through behavioral testing, capability verification on smaller datasets, architectural validation at reduced scales, and comprehensive process documentation.

In response, the research community has developed standardized benchmarks and evaluation frameworks that validate model capabilities without requiring access to implementation details. Platforms like the Open LLM Leaderboard by \cite{huggingface2023leaderboard}, HELM by \cite{liang2022holistic}, MMLU by \cite{hendrycks2020measuring}, and Chatbot Arena by \cite{chiang2024chatbot} represent conceptual replicability approaches, as they test the central hypotheses about model capabilities through experimental designs different from those used during development.

This case demonstrates the critical role of conceptual replicability in modern AI validation. When other validation approaches become technically or economically infeasible, conceptual replicability provides a path for maintaining scientific rigor by ensuring claims about AI capabilities remain testable and falsifiable, even when working with systems whose inner workings remain inaccessible.

\section{Limitations and Boundaries of the Framework}
While our reproducibility framework provides structured guidance for validation in AI/ML research, it has important limitations regarding its scope and application.

\subsection{Limited Applicability to Non-Empirical Research}
Our framework is designed primarily for empirical, data-centric AI/ML research and has limited applicability to theoretical work (focused on mathematical proofs), algorithmic innovations without extensive experimental validation, hardware-specific research requiring proprietary equipment, and AI ethics or philosophical inquiries. For these categories, different validation approaches would be necessary.

\subsection{Beyond Correctness: What Validation Does Not Address}
Our framework validates correctness, robustness, and generalizability but does not address several critical dimensions of AI/ML research: fairness and equity across demographic groups, ethical implications including potential misuse, broader societal consequences, and alignment with human values. These considerations require complementary frameworks beyond traditional scientific validation. Reproducibility is necessary but insufficient for responsible AI/ML research.

\subsection{Broader Reward Systems and Incentives}
While our framework clarifies definitions, terminological confusion is only one factor in the reproducibility crisis. Broader academic and industry reward systems significantly influence reproducibility practices, including publication incentives favoring novelty, resource constraints discouraging thorough documentation, and commercial considerations limiting data sharing. Addressing these systemic challenges is beyond our scope but essential for comprehensive resolution of the reproducibility crisis.

\section{Conclusions}
In response to the challenges posed by the reproducibility crisis within the artificial intelligence and machine learning domain, our work has established a clear framework and definitions for critical terms such as repeatability, reproducibility, and replicability. By differentiating between these key concepts and exploring their implications for scientific rigor and reliability, we aim to clarify their meanings and underline their vital role in enhancing trust in scientific research. Ultimately, our goal is to foster a scientific environment where findings are not only repeatable and reproducible but also robustly replicable across various contexts, thereby advancing the field with trustworthy and verifiable knowledge.

\balance
\bibliographystyle{chicago}
\bibliography{references}
\end{document}